# Optical, magneto-optical properties and fiber-drawing ability of tellurite glasses in the TeO$_2$–ZnO–BaO ternary system

J. Hrabovsky[1,2*], L. Strizik[3], F. Desevedavy[4], S. Tazlaru[1], M. Kucera[1], L. Nowak[1], R. Krystufek[5,6], J. Mistrik[7,8], V. Dedic[1], V. Kopecky Jr.[1], G. Gadret[4], T. Wagner[3,7], F. Smektala[4], M. Veis[1]

[1] Faculty of Mathematics and Physics, Charles University, Ke Karlovu 2027/3, Prague 2, 121 16, Czech Republic

[2] HiLASE Centre, Institute of Physics of the Czech Academy of Sciences, Za Radnicí 828, Dolní Břežany, 252 41, Czech Republic

[3] Department of General and Inorganic Chemistry, Faculty of Chemical Technology, University of Pardubice, Studentská 95, Pardubice 532 10, Czech Republic

[4] ICB Laboratoire Interdisciplinaire Carnot Bourgogne, UMR 6303 CNRS-Université de Bourgogne Franche Comte, 9 Av. Alain Savary, 21078 Dijon, France.

[5] Institute of Organic Chemistry and Biochemistry of the Czech Academy of Sciences, Prague 6 16610, Czech Republic

[6] Department of Physical and Macromolecular Chemistry, Faculty of Science, Charles University, Prague 2 128 43, Czech Republic

[7] Center of Materials and Nanotechnologies, Faculty of Chemical Technology, University of Pardubice, nám. Čs. Legií 565, Pardubice 530 02, Czech Republic

[8] Institute of Applied Physics and Mathematics, Faculty of Chemical Technology, University of Pardubice, Studentská 95, Pardubice 532 10, Czech Republic

*The corresponding authors e-mail: hrabovj@karlov.mff.cuni.cz

**ABSTRACT**

The presented work is focused on the optical and magneto-optical characterization of $TeO_2$-$ZnO$-$BaO$ (TZB) tellurite glasses. We investigated the refractive index and extinction coefficient dispersion by spectroscopic ellipsometry from ultraviolet, $\lambda \approx 0.193$ μm, up to mid-infrared, $\lambda \approx 25$ μm spectral region. Studied glasses exhibited large values of linear ($n_{632} \approx 1.91$–$2.09$) and non-linear refractive index ($n_2 \approx 1.20$–$2.67 \times 10^{-11}$ esu), Verdet constant ($V_{632} \approx 22$–$33$ radT$^{-1}$m$^{-1}$) and optical band gap energy ($E_g \approx 3.7$–$4.1$ eV). The materials characterization revealed that BaO substitution by ZnO leads (at constant content of $TeO_2$) to an increase in linear and nonlinear refractive index as well as Verdet constant while the optical band gap energy decreases. Fiber drawing ability of $TeO_2$-$ZnO$-$BaO$ glassy system has been demonstrated on $60TeO_2$–$20ZnO$–$20BaO$ glass with presented mid-infrared attenuation coefficient. Specific parameters such as dispersion and single oscillator energy, Abbe number, and first-/third-order optical susceptibility are enclosed together with the values of magneto-optic anomaly derived from the calculation of measured dispersion of the refractive index.

**ABBREVIATIONS**

MIR – mid-infrared; MO – magneto-optical; PA – phonon absorption; PCR– Prism coupled refractometry; SE – spectroscopic ellipsometry; T-L – Tauc-Lorentz; TZB – $TeO_2$–$ZnO$–$BaO$; UV – ultraviolet

1. Introduction

Tellurite glasses have gained significant attention in recent years due to their unique optical and magneto-optical (MO) properties[1,2]. In general, tellurite glasses have high refractive index, dielectric constant and Raman gain coefficient, low phonon energies and wide transparency window from ultraviolet (UV) to the mid-infrared (MIR) spectral region, making them promising for optical applications such as fiber optics, amplifiers, and lasers[1–6]. The other advantage of tellurite glasses is their good glass-forming ability[7] and relatively easy preparation at mild temperatures, usually <1000 °C[1]. However, it should be noted that the preparation of pure $TeO_2$ glass by melt-quenching technique is still quite challenging despite the recent progress in the use of the intermittent quenching technique [8]. Therefore, the addition of even a small amount of glass-forming oxides ($P_2O_5$ and $B_2O_3$) or/and another components to the $TeO_2$ such as alkaline, alkaline-earth and heavy-metal oxides or halides do not only promote a glass-forming ability but also allow tunability of resulting physico-chemical properties [1,2]. Despite the promising optical and magneto-optical properties of tellurite glass, a limited number of studies is devoted to determination of refractive index dispersion from ultraviolet (UV) up to MIR spectral region [1,2,9,10] and MO properties, such as Faraday rotation. Many undoped tellurite glasses belong to the group of diamagnetic glasses [11] with generally presented values of Verdet constant at 632 nm $V_{632} \approx 27$ radT$^{-1}$m$^{-1}$ [10], lower compared to crystals or paramagnetic glasses [12,13] but higher than other light flint glasses[10]. Investigation of MO properties in tellurite glasses has been studied at limited wavelengths range and compositions in binary, ternary and multicomponent tellurite glasses [14–20].

Recently, we have studied the glass-forming ability and basic physico-chemical properties of $TeO_2$-ZnO-BaO (TZB) ternary-system glasses [21]. In present work we extended our study[21] by investigating the dispersion of the refractive index of selected TZB glasses from ultraviolet up to mid-infrared spectral region ($\lambda \approx 193$ nm– 25 µm) with an estimation of their non-linear refractive index by semi-empirical rule in accordance with Ref. [22], optical band gap energy $E_g$, and values of Verdet constant. Moreover, we demonstrate that the studied TZB glasses may be utilized in fiber optics by drawing of multimode optical fiber. The obtained results show that glasses from TZB group could be used for development of magneto-active or non-linear optical devices.

## 2. Experimental procedure
### 2.1. Bulk glass and optical fiber preform preparation

The $TeO_2$–ZnO–BaO glasses have been prepared using the conventional melt-quenching technique from high-purity compounds of $TeO_2$ (Alfa Aesar, 4N), $BaCO_3$ (Alfa Aesar, 4N) and ZnO (Alfa Aesar, 4N). Our investigation is focused on nine bulk samples of the chemical compositions shown in Fig. 1 with denoted glass-forming region. Detailed conditions of preparation of glasses, verification of their amorphous nature by X-ray diffraction and chemical composition by energy-dispersive X-ray spectroscopy are provided in Ref. [21]. The $(TeO_2)_{60}(ZnO)_{20}(BaO)_{20}$ glassy preform for optical fiber drawing has been made at room mosphere without any purification process by melting the starting compounds $TeO_2$ (technical grade), ZnO (4N), $BaCO_3$ (4N) in Pt crucible at 900 °C for 25 min with a total batch weight of 25 g. The melt was poured into a preheated (≈270 °C) cylindrical brass mould with an inner diameter of ≈9 mm and a length of cylinder ≈80 mm. Glasses have been annealed for 8 h near the glass transition temperature $T_g$ to relax the internal strain

The fiber drawing from a glassy preform was performed by using a dedicated 3 m high optical fiber draw tower. The preform was slowly fed into the furnace under an He gas flow (3 L min$^{-1}$) and the temperature was gradually increased to its softening temperature. The glass was brought to its softening temperature regime while the drawing parameters (drum rotation, preform feed, tension and temperature) were continuously monitored to produce the optical fibers of the targeted diameter. Optical characterization was carried out on plane-parallel, one-side polished glassy blocks of thickness ≈0.5 mm. Dispersion of refractive index and extinction coefficient measurements were performed using two variable-angle spectroscopic ellipsometers. The first, variable angle Mueller matrix spectroscopic ellipsometer RC2 (J.A. Woollam Co., Inc.), has been used for the measurement of optical constants (refractive index $n$ and extinction coefficient $k$) in the UV-VIS-NIR spectral range of 0.193–1.690 μm (6.42–0.73 eV) with a step of ≈1 nm at angles of incidence 60°, 65° and 70°. The second, rotating compensator IR-VASE® (J. A. Woollam Co., Inc.) ellipsometer has been used for determination of optical constants in the spectral IR region of 1.7–25 μm (25 scans/wavenumber, 15 spectra per revolution with a wavenumber step of ≈ 8 cm$^{-1}$) at the same angles of light incidence. Near normal incidence optical reflectance and normal optical transmittance were measured by the same instruments. WVASE32 software was used for modelling of the combined set of measured data from both ellipsometers. Ellipsometric parameters $\Psi$ and $\Delta$ were fitted using multi-layer spectroscopic ellipsometry (SE) model including layer and surface roughness[23,24] to increase both the sensitivity and accuracy of the determination of short- and long-wavelength absorption edge of measured samples. Room temperature Faraday MO spectra were collected by a spectrometer based on the compensation method and utilizing a rotating polarizer approach for modulation and synchronous detection in the wavelength range of 500–2400 nm (2.5–0.52 eV). The light from a highly stabilized bulb (IR) and Xe lamp (UV-Vis) was monochromated, passed through the sample at normal incidence, then detected by a photomultiplier PbS detector. Applied magnetic field from an electromagnet parallel to the optical beam and perpendicular to the sample surface was set to 0.3 T. Faraday hysteresis loops were measured at two wavelengths of 550 nm and 1550 nm varying an applied magnetic field up to 0.75 T to verify the diamagnetic nature of prepared glasses. The experimental setup for the measurement of Faraday rotation is shown in Fig. S1. The fiber transmission losses were measured using the cutback method on several meter lengths of fiber with a Nicolet 6700 Fourier-transform Infrared (FTIR) spectrometer in the 1.0–4.5 μm range. For the measurement, a halogen lamp emitting from 0.1 to 4.5 μm was used as an optical source. The fiber output power was detected using an InSb photodetector and the fiber input and output were manually cleaved.

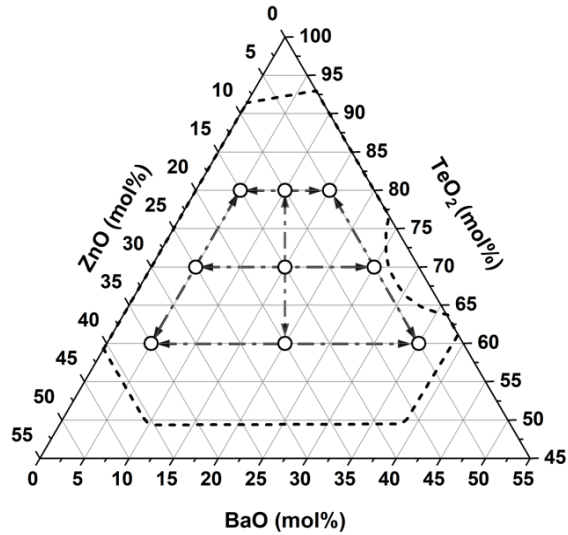

**Fig.1.** TeO$_2$–ZnO–BaO ternary diagram with indicated chemical compositions of studied glasses (black circles), and the glass-forming boundary (dashed line) [7,21]. Shown connections (dashdotted arrowed lines) represent the concentration trends with constant content of TeO$_2$ (horizontal lines), ZnO (left inclined) BaO (right inclined) or the equimolar substitution of TeO$_2$ by both ZnO and BaO (vertical line).

## 3. Results and discussion

Selected samples of the TZB glasses prepared in previous work [21] have been chosen to study the linear and nonlinear optical and MO properties with respect to chemical composition (listed in Table 1). The amorphous nature of prepared samples was verified by X-ray diffraction analysis with no detected crystallization peaks (Fig. S2) [21]. Calculated values of molar volume ($V_M$) are listed in Table 2 (taken from Ref. [21]).

**Table 1** Chemical compositions of TeO$_2$–ZnO–BaO glasses with samples ID.

| Sample ID | TeO$_2$ | ZnO | BaO |
|---|---|---|---|
| | (mol. %) | | |
| T80Z15B5 | 80 | 15 | 5 |
| T80Z10B10 | 80 | 10 | 10 |
| T80Z5B15 | 80 | 5 | 15 |
| T70Z25B5 | 70 | 25 | 5 |
| T70Z15B15 | 70 | 15 | 15 |
| T70Z5B25 | 70 | 5 | 25 |
| T60Z35B5 | 60 | 35 | 5 |
| T60Z20B20 | 60 | 20 | 20 |
| T60Z5B35 | 60 | 5 | 35 |

## 3.1. Linear optical properties

Optical constants of studied glasses as a function of wavelength over a wide spectral range ($\lambda \approx 193\text{–}25\,000$ nm) determined by spectroscopic ellipsometry are presented in Fig. 2 and are compared to values of $n$ measured by the Prism coupled refractometry (PCR) method at wavelengths of 632 ($n_{632}$), 1064 ($n_{1064}$) and 1550 nm ($n_{1550}$) [21]. Spectral dependence of refractive index $n$ and extinction coefficient $k$ was modelled using Tauc-Lorentz (T-L) oscillator (UV part) and 5 Gaussian oscillators (IR part) whose parameters were obtained by the minimalization of mean square error between experiment and model data. Tauc-Lorentz oscillator is frequently used for modelling the short wavelength absorption edge of amorphous/glassy materials [25,26], whereas the sum of several Gauss and/or Lorentz oscillators was reported for the description of phononic/long-wavelength absorption edge [27,28].

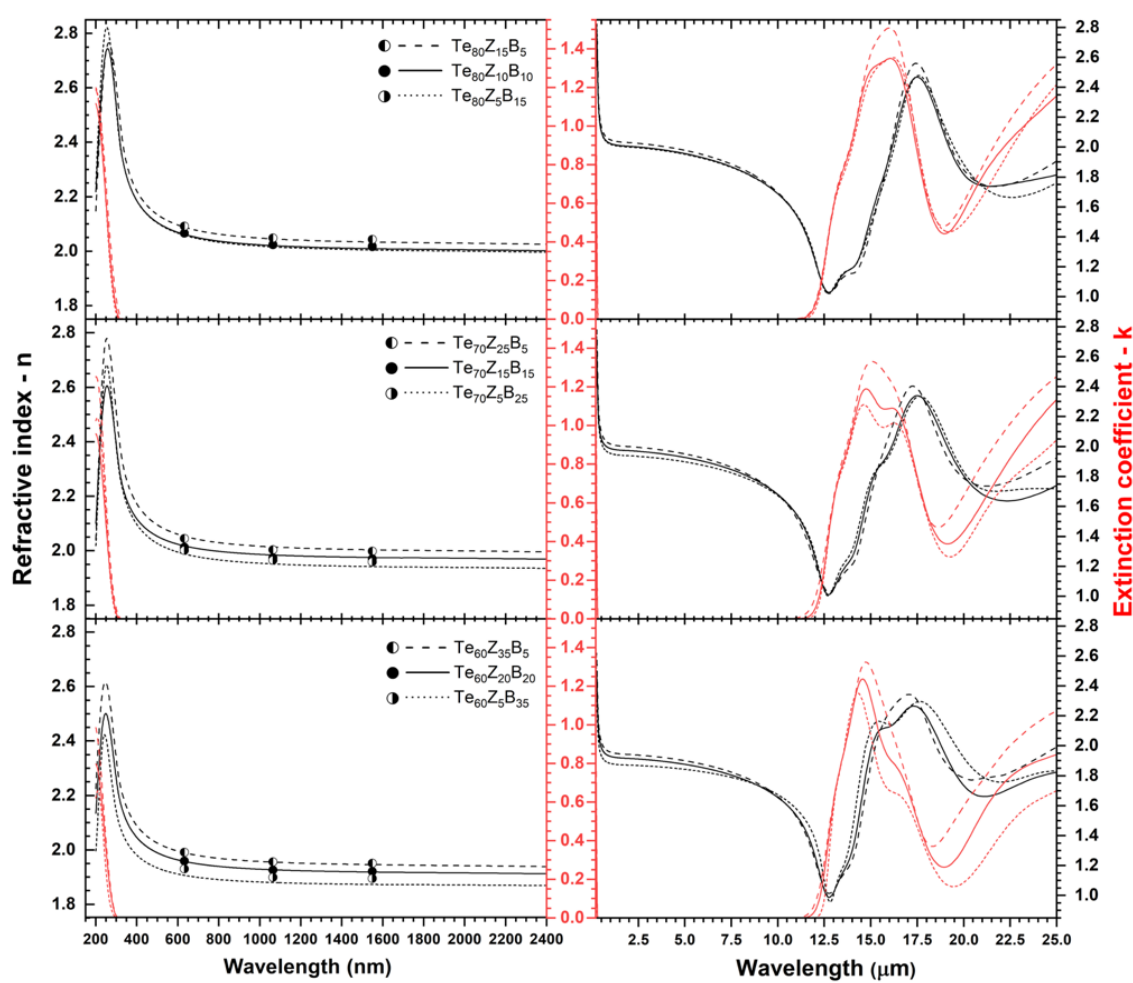

**Fig.2.** Spectral dispersion of refractive index (black lines) and extinction coefficient (red lines) of TeO$_2$–ZnO–BaO glasses. Black-filled and half-filled circles represent $n$ values obtained via Prism coupled refractometry method at $\lambda \approx 633$, 1064 and 1550 nm in Ref. [21].

The derived optical parameters are summarized in Table 2 with regard to the chemical compositions of TZB glasses and follow the concentration lines with constant $TeO_2$, ZnO and BaO content. Both the refractive index and extinction coefficient for all glasses are enclosed in the Supplementary information in the whole observed spectral range. From Table 2 it is visible, that the values of the refractive index reach $n_{632}$ = 1.91–2.09 and $n_{1550}$ = 1.87–2.03 respectively. Those are lower value than for pure $TeO_2$ glass ($n_{632}$ = 2.184) [29], but higher compared to the commonly used $TeO_2$–ZnO–$Na_2O$ glasses or binary $TeO_2$–BaO glasses[1]. The combination of both wide concentration variability and the sensitive dependence of the refractive index on composition makes these glasses an interesting material for precise chromatic dispersion engineering.

The observed deviation between the experimental values of refractive index determined by PCR method and SE in Fig.2 are below ±0.01 for most samples, except the samples (T60Z5B35 and T70Z5B25) with observed ±0.02 deviation. This deviation can result from different polishing processes or/and thickness reduction as the previous measurements were carried out on hand-polished ≈4 mm thick blocks compared to the ≈0.5 mm semiautomatically polished samples presented in this work. Our spectra also follow previous results made on T60Z20B20 [30] glass composition, where observed deviation was found to be ±0.01 for both presented wavelengths of 632 and 1550 nm. We can thus conclude good sample preparation reproducibility within the TZB glassy system. Optical material dispersion was derived using the known values of refractive index at Franhofer spectral lines at $\lambda \approx$ 656.3, 587.6 and 486.1 nm. Obtained values of Abbe number $v_D = (n_{587.6} - 1)/(n_{486.1} - n_{656.3})$ lie within a region of 18.5 to 26.3 (higher than for binary $TeO_2$–ZnO/BaO glasses[1]) which manifest that TZB glasses are dense flint glasses.

A detailed explanation of composition dependent refractive index evolution was presented in Ref [21] and can be explained on the basis of the observed change in density and molar volume ($V_M$) or assigned with the decay of glass structure and resulting transformation of $TeO_4$ ($\Lambda_{TeO4}$ = 0.99) to $TeO_3$ ($\Lambda_{TeO3}$ = 0.82) structural units associated with a change in the optical basicity ($\Lambda$) of the glasses [31–33].

The optical band gap energy of studied glasses has been determined by the fit of SE data to the T-L oscillator and the obtained parameters as well as the description of T-L function are enclosed in the Supplementary information (Table S1). The energies were found to be in the interval of $E_g \approx$ 3.73–4.06 eV (Table 2). It is clearly observed that substitution of $TeO_2$ or ZnO by BaO increases the $E_g$ value. The same behavior is observed when $TeO_2$ is substituted by ZnO at constant BaO content. Due to inconsistencies in optical band gap determination within the literature, such as different method of preparation, thickness of samples and method used, it is difficult to directly compare the obtained values, even with the binary $TeO_2$-based glasses. However, Mallawany [1] refers to an observable increase in optical band gap energy for both binary TeO$_2$–ZnO/BaO systems compared to the estimated optical band gap energy of pure TeO$_2$ glass $E_g$ = 3.37 eV (220 μm thick sample, estimated from optical absorption spectrum) [29]. Composition dependence of both, the refractive index and energy band gap, is illustrated in Fig.3 and exhibits opposite trends in accordance with Moss rule[34].

Considering all derived optical parameters in Table 2, we can conclude that substitution of TeO$_2$ by ZnO or BaO in TZB leads to a decrease of refractive index $n$, molar refractivity $R_M$ (Eq.1a) and polarizability $\alpha$ (Eq.1b, where N$_A$ is the Avogadro's number) and an increase of optical band gap energy $E_g$. However, substitution of TeO$_2$ by ZnO (at constant BaO) reduces a molar volume $V_M$ of glass while substitution of TeO$_2$ by BaO (at constant ZnO content) leads to increment of molar volume $V_M$. Similarly, the substitution of ZnO by BaO (at constant TeO$_2$ content) leads to decrease of refractive index value and an increase of $V_M$, $R_M$ as well as electronic polarizability [21].

$$(a) \quad R_M = \left(\frac{n^2-1}{n^2+2}\right) \times V_M \qquad (b) \quad \alpha = \left(\frac{3}{4\pi N_A}\right) \times R_M \qquad (1)$$

**Table 2.** Optical parameters of TeO$_2$–ZnO–BaO glasses: refractive index ($n_\lambda$) at wavelength of $\lambda \approx$ 632 or 1550 nm, optical band gap energy ($E_g$) value, Abbe number ($\nu_D$), molar refractivity ($R_M$), electronic polarizability ($\alpha$), molar volume ($V_M$) from [14] and calculated ratios $\alpha/V_M$.

| Sample ID | $n_{632}$ | $n_{1550}$ | $E_g$ (eV) | $\nu_D$ | $R_{M,1550}$ (cm$^3$mol$^{-1}$) | $\alpha$ (Å$^3$) | $V_M$ (cm$^3$mol$^{-1}$) | $\alpha/V_M$ (×10$^{-25}$ mol) |
|---|---|---|---|---|---|---|---|---|
| T80Z15B5  | 2.088 | 2.034 | 3.73 | 18.5 | 13.67 | 5.42 | 26.81 | 2.02 |
| T80Z10B10 | 2.062 | 2.009 | 3.86 | 19.1 | 13.82 | 5.48 | 27.45 | 2.00 |
| T80Z5B15  | 2.059 | 2.004 | 3.94 | 18.5 | 14.07 | 5.58 | 28.14 | 1.98 |
| T70Z25B5  | 2.053 | 2.003 | 3.79 | 19.8 | 12.82 | 5.08 | 25.63 | 1.98 |
| T70Z15B15 | 2.018 | 1.975 | 3.89 | 22.7 | 13.31 | 5.28 | 26.98 | 1.96 |
| T70Z5B25  | 1.988 | 1.942 | 3.95 | 20.1 | 13.59 | 5.39 | 28.35 | 1.90 |
| T60Z35B5  | 1.989 | 1.946 | 3.99 | 21.8 | 11.79 | 4.67 | 24.41 | 1.91 |
| T60Z20B20 | 1.958 | 1.919 | 3.99 | 23.7 | 12.52 | 4.96 | 26.51 | 1.87 |
| T60Z5B35  | 1.907 | 1.873 | 4.06 | 26.3 | 13.15 | 5.21 | 28.94 | 1.80 |

In an insulating crystalline solid, the light absorption at the infrared edge is principally caused by harmonic and anharmonic terms in the lattice potential leading to fundamental and higher harmonics of the lattice resonances, respectively. This phenomenon is usually referred to as multiphonon absorption. As structural disorder increases, these optical modes broaden, and additional ones appear. Therefore, multiphonon absorption in noncrystalline (amorphous) materials, such as glasses, is represented by significantly broader oscillators than those encountered in crystals [35]. The SE data in the infrared spectral region were modelled by sum of 5 Gaussian oscillators from which we obtained position of long-wavelength/phonon absorption (PA) edge of studied TZB glasses. Although the general methodology for long-wavelength/phonon absorption edge determination still needs to be fully developed for non-crystalline solids, two original approaches are proposed here. In the first case, the deduced position of PA1 edge has been determined from the linear plot of $\alpha_{abs}$ as the function of wavenumber in cm$^{-1}$, where $\alpha_{abs}$ is the absorption coefficient derived from optical model fit (Fig. S3). Derived position of PA1 lies in the region of wavenumbers from 833–931 cm$^{-1}$ corresponding to wavelengths of 12–10.7 μm (Table 3). It is evident that the substitution of ZnO by BaO leads to a red shift of this absorption band which may be connected to different bond energies or due to the higher mass of Ba atoms compared to Zn. However, this method gives significantly higher values of obtained wavelength positions of PA edge compared to FTIR measurements presented in our previous work (~6 μm)[21]. Therefore, the infrared absorption coefficient of material was characterized in the second case by an exponential function in the form[35]

$$\alpha_{abs} = \alpha_0 \, exp\left(-\gamma_{abs}\frac{\omega}{\omega_0}\right)$$

(2)

where $\alpha_0$ is a constant (same dimension as the absorption coefficient), $\gamma_{abs}$ is a dimensionless constant (typically close to be 4), $\omega_0$ is the parameter of the maximum frequency and $\omega$ is the wavenumber. This approximation of PA2 absorption edge was found to be valid for the range of absorption coefficients from 0.001 to 10 cm$^{-1}$. However, within this abs. coeff. range we observed two distinct regimes with different linear slopes in semilogarithmic plots of $\alpha_{abs}$ as the function of wavenumber. Transition between these regimes occurs in region of approximately 0.1–1.0 cm$^{-1}$ (Figure S4) and other non-linear behavior in the range under $1\times10^{-4}$ cm$^{-1}$ follows. Because of that, the experimental data of absorption coefficient were fitted using Eq.2 in abs. coeff. range 0.1 to 0.001 cm$^{-1}$ (lower limit of method) for all TZB glasses (Figure S5) and further extrapolated for lower values of absorption coefficient. Note, that the disadvantage of using the exponential approximation method is the difficulty of determining the zero value of absorption coefficient in semilogarithmic plots. The derived positions of PA2 edge are therefore listed in Supplementary information (Table S2) for various "zero" values of absorption coefficient ranging from $1\times10^{-3}$ cm$^{-1}$ to $1\times10^{-6}$ cm$^{-1}$. The value of $\alpha_{abs} = 1\times10^{-4}$ cm$^{-1}$ was then used to determine the position of PA2 due to observed change between linear and nonlinear behavior of $\alpha_{abs}$ in this region. Derived position of PA2 lies in the region of wavenumbers from 922–1406 cm$^{-1}$ corresponding to wavelengths of 6.9–10.8 μm (Table 3), which provides a better agreement with the experimental data from FTIR[21] compared to the first method of linear PA1 determination. However, it should be noted that this is only a rough estimation as the methodology for evaluating of the position of long-wavelength/phonon absorption edge in amorphous materials is not entirely clear and for practical applications utilizing bulk samples, *i.e.* thick samples, the transmittance in the infrared part of the spectrum is high up to ~6 μm due to Beer-Lambert law.

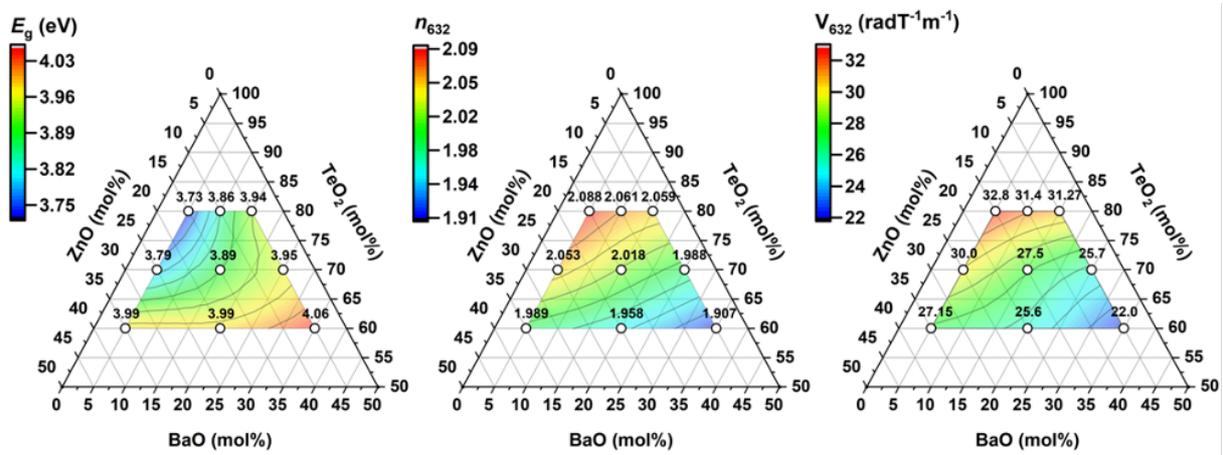

**Fig.3** Evolution of the optical band gap ($E_g$), refractive index ($n_{632}$) and Verdet constant ($V_{632}$) determined at 632 nm with respect to chemical composition of TeO$_2$–ZnO–BaO glasses.

### 3.2. Non-linear optical properties

Spectral dependence of refractive index in the transparent range has been also used for determination of dispersion energy ($E_d$) and single oscillator energy ($E_0$) utilizing the single-oscillator approximation according to Wemple and DiDomenico (WDD)[36]:

$$n^2 - 1 = \frac{E_0 E_d}{E_0^2 - E^2} \quad (3)$$

The intercept and slope of the linear fit to $(n^2 - 1)^{-1}$ versus square of photon energy $E^2$ dependency (see Fig. S6) in transparent spectral region of 0.73–2.48 eV (λ ≈ 500–1690 nm) provides parameters $E_d$, $E_0$ and refractive index $n(0)$ for photon energy $E \to 0$ eV (Eq. 3) as is presented in Table 3. The spectral region $E \approx 0.73$–2.48 eV (500–

1690 nm) for the obtaining the $E_0$ and the $E_d$ parameters has been chosen to be fairly below the optical band gap energy (interband transitions or excitonic absorption) and above the phonon absorption. Both the ultraviolet (interband transitions) as well as the infrared (phonon modes) absorption contributes to the refractive index dispersion and eventually leads to nonlinearity of the assumed linear $(n^2 - 1)^{-1}$ versus $E^2$ dependency. Although the nonlinear dependency may be quite well fitted by three-pole Sellmeier model, the linking of this model to the chemical structure is not fully clear. Thus, the simple single-oscillator WDD model has been used in selected spectral region to extract structural features related to dispersion energy $E_d$. The dispersion energy $E_d$ of studied TZB glasses obtained from fit of Eq. 3 to data in Fig. S6 is corelated to the structure of studied materials via the empirical formula $E_d = \beta\, N_c\, N_e\, Z_a$ [22], where $N_c$ is the coordination number of the cation nearest neighbour to the anion, $N_e$ is the effective total number of valence electrons (cores excluded) per anion, $Z_a$ is the formal chemical valency of the anion and $\beta$ is empirical parameter describing the polarity of chemical bonding; $\beta \approx (0.37 \pm 0.04)$ eV for covalently bonded materials and $\beta \approx (0.26 \pm 0.03)$ eV for ionic materials. Unfortunately, the reduction of $E_d$ for studied ternary TZB glasses seems to be complicated, since no deep information on the atomic (dis)order of TZB glasses is known. Obtained parameters $E_0$, $E_d$ were used for calculation of refractive index value for photon energy $E \to 0$, i.e. $n(0) = (1 + E_d/E_0)^{1/2}$ and subsequently for estimation of the first-order $\chi^{(1)}$ and third-order $\chi^{(3)}$ optical susceptibility and nonlinear refractive index values $n_2$ in esu units according to the method suggested by Tichá and Tichý [22] including the recommendation by Górski et al.[37]. For comparison, we also calculated values of $n_2$ from $E_d$ and $E_0$ in Refs. [22,36,38–41] for the other materials which are listed in Table 4. The nonlinear refractive index of studied TZB glasses is higher approximately by ~1.5 orders of magnitude than those for fused $SiO_2$ and approximately by ~1.3 orders of magnitude than nonlinear KHP ($KH_2PO_4$) crystal. Moreover, it may be seen that nonlinear refractive index value increases by substitution of ZnO for BaO as well as by increasing of the overall $TeO_2$ concentration in TZB glasses. Therefore, TZB glasses may be promising materials in nonlinear optics, however, the true figure of merit must be experimentally determined [42].

**Table 3.** Optical parameters of $TeO_2$–ZnO–BaO glasses: measured refractive index ($n_{1550}$), position of photonic/long wavelength absorption edge (PA) and band gap energy ($E_g$) and calculated values of the non-linear refractive index ($n_2$), refractive index ($n(0)$) for photon energies $E \to 0$ eV, dispersion energy ($E_d$), single oscillator energy ($E_0$) and first-order ($\chi^{(1)}$) and third-order ($\chi^{(3)}$) optical susceptibility.

| Sample ID | $n_{1550}$ | PA1 (cm$^{-1}$) | PA2 (cm$^{-1}$) | $E_g$ (eV) | $E_d$ (eV) | $E_0$ (eV) | $n(0)$ | $\chi^{(1)}$ (esu) | $\chi^{(3)}$ ($10^{-13}$ esu) | $n_2$ ($10^{-11}$ esu) |
|---|---|---|---|---|---|---|---|---|---|---|
| T80Z15B5 | 2.034 | 911 | 1406 | 3.73 | 21.39 | 6.92 | 2.02 | 0.246 | 6.23 | 2.67 |
| T80Z10B10 | 2.009 | 892 | 1033 | 3.86 | 20.96 | 7.00 | 2.00 | 0.238 | 5.49 | 2.38 |
| T80Z5B15 | 2.004 | 877 | 1326 | 3.94 | 20.46 | 6.88 | 1.99 | 0.237 | 5.34 | 2.32 |
| T70Z25B5 | 2.003 | 931 | 1451 | 3.79 | 21.25 | 7.15 | 1.99 | 0.237 | 5.34 | 2.32 |
| T70Z15B15 | 1.975 | 885 | 1338 | 3.89 | 21.82 | 7.61 | 1.97 | 0.228 | 4.61 | 2.03 |
| T70Z5B25 | 1.942 | 862 | 970 | 3.95 | 19.67 | 7.19 | 1.93 | 0.218 | 3.81 | 1.71 |
| T60Z35B5 | 1.946 | 902 | 1276 | 3.99 | 20.49 | 7.44 | 1.94 | 0.219 | 3.92 | 1.75 |
| T60Z20B20 | 1.919 | 874 | 1171 | 3.99 | 20.55 | 7.75 | 1.91 | 0.211 | 3.37 | 1.53 |
| T60Z5B35 | 1.873 | 833 | 922 | 4.06 | 20.22 | 8.14 | 1.87 | 0.198 | 2.59 | 1.20 |

**Table 4** Optical parameters of selected reference materials calculated using values taken from Ref.[22,36,38–41]: dispersion energy ($E_d$), single oscillator energy ($E_0$), non-linear refractive index ($n_2$), refractive index ($n(0)$) for photon energies $E \rightarrow 0$ eV, first-order ($\chi^{(1)}$) and third-order ($\chi^{(3)}$) optical susceptibility.

| Material | $E_d$ | $E_0$ | $n(0)$ | $\chi^{(1)}$ | $\chi^{(3)}$ | $n_2$ |
|---|---|---|---|---|---|---|
| | (eV) | | | | (esu) | |
| $g$-SiO$_2$ | 14.7 | 13.4 | 1.45 | 0.087 | $9.87 \times 10^{-15}$ | $5.91 \times 10^{-13}$ |
| $c$-SiO$_2$ | 18.1 | 13.3 | 1.54 | 0.108 | $2.34 \times 10^{-14}$ | $1.32 \times 10^{-12}$ |
| $\alpha$-TeO$_2$ ($o$) | 23.5 | 6.33 | 2.17 | 0.295 | $1.29 \times 10^{-12}$ | $5.17 \times 10^{-11}$ |
| $\alpha$-TeO$_2$ ($e$) | 26.2 | 6.08 | 2.30 | 0.343 | $2.35 \times 10^{-12}$ | $8.85 \times 10^{-11}$ |
| $c$-ZnO | 16.3 | 6.1 | 1.92 | 0.213 | $3.48 \times 10^{-13}$ | $1.57 \times 10^{-11}$ |
| $c$-BaO | 19 | 7.1 | 1.92 | 0.213 | $3.50 \times 10^{-13}$ | $1.58 \times 10^{-11}$ |
| As$_2$S$_3$ | 25.2 | 6.0 | 2.28 | 0.334 | $2.12 \times 10^{-12}$ | $8.07 \times 10^{-11}$ |
| As$_2$Se$_3$ | 28.0 | 4.7 | 2.64 | 0.474 | $8.59 \times 10^{-12}$ | $2.82 \times 10^{-10}$ |
| As$_2$Te$_3$ | 28.05 | 2.325 | 3.61 | 0.960 | $1.44 \times 10^{-10}$ | $3.46 \times 10^{-9}$ |
| KH$_2$PO$_4$ | 16 | 12.8 | 1.50 | 0.099 | $1.66 \times 10^{-14}$ | $9.62 \times 10^{-13}$ |
| LiNbO$_3$ | 25.9 | 6.65 | 2.21 | 0.310 | $1.57 \times 10^{-12}$ | $6.15 \times 10^{-11}$ |

### 3.3. Magneto-optical properties, Verdet constant

The angle of the Faraday rotation $\theta_F$ of diamagnetic or paramagnetic materials is defined as $\theta_F = VBl$, where $B$ is the magnetic field applied in the direction of propagation of light, $l$ is sample thickness, and $V$ is the Verdet constant in rad T$^{-1}$ m$^{-1}$. With a known magnitude of magnetic field and sample thickness, the value of the Verdet constant can be easily determined. The spectral dependencies of Verdet constant deduced from $\theta_F$ measurements are shown in Fig.4 for all samples.

Moreover, the experimental values measured at 632 nm are listed in Table 5 and vary from 22.0 to 32.8 rad T$^{-1}$ m$^{-1}$ (0.076–0.113 min Oe$^{-1}$ cm$^{-1}$), and follow the same compositional dependence as the refractive index (Fig. 3). Obtained values are therefore sufficient for construction of optical isolators (with $\theta = 45°$) using optical fiber with the length in the range of 24 to 36 mm and an electromagnet, $B = 1$ T or longer fiber length of 36 to 54 mm with a permanent neodymium magnet, 0.5 T. Verdet constants for TeO$_2$ rich compositions are then higher approximately by one order of magnitude than those for fused SiO$_2$ (633 nm, 3.7 rad T$^{-1}$ m$^{-1}$) [43]. Glasses from TZB system also exhibit greater Faraday rotation than most of the previously reported binary TeZn (≈25 rad T$^{-1}$ m$^{-1}$)[14], ternary TeZnNa (27.1/28.1 rad·T$^{-1}$m$^{-1}$)[15,17], TeZnLa (≈29 rad T$^{-1}$ m$^{-1}$) or some other multicomponent TeZnNaBa (26.5 rad T$^{-1}$ m$^{-1}$)[18], TeZnPbNbLa (29.7 rad T$^{-1}$ m$^{-1}$)[19], TeWPbLa (600 nm, ≈19 rad T$^{-1}$ m$^{-1}$)[20], TeZnAlP (≈5 rad T$^{-1}$ m$^{-1}$)[44] tellurite glasses without additional rare earth ion content (unlabeled compositions were measured at 632 nm). Overall, the Verdet constant of tellurite glasses is usually reported as ≈27 rad T$^{-1}$ m$^{-1}$, lower compared to crystals or paramagnetic glasses [12,13] but higher than other light flint glasses[10]. Experimental values of Verdet constant measured at 1550 nm are also enclosed in Table 5 and vary from 3.3 to 4.7 rad T$^{-1}$ m$^{-1}$ what are greater values than for TeLaW glasses (2.75 to 3.95 rad T$^{-1}$ m$^{-1}$)[45].

**Table 5** Obtained optical and magneto-optical parameters of TeO$_2$–ZnO–BaO glasses: experimental Verdet constant ($V$) at 632 nm and 1550 nm and calculated Verdet constant ($V_{632,cal}$) at 632 nm (Eq.5), magneto-optic anomaly ($\gamma$), Sellmeier parameters for description of spectral dependence of Verdet constant ($A_M$, $B_M$) (Eq.4a) and refractive index ($B_1$, $C_1$, $B_2$, $C_2$) (Eq.4b) for transparent region of used glasses. Numbers in parentheses represents the tolerance on the last digit.

| Sample ID | $V_{632}$ | $V_{632,cal}$ | $V_{1550}$ | $\gamma$ | $A_M$ | $B_M$ | $B_1$ | $C_1$ | $B_2$ | $C_2$ |
| --- | --- | --- | --- | --- | --- | --- | --- | --- | --- | --- |
| | (rad T$^{-1}$ m$^{-1}$) | | | | (rad T$^{-1}$ m$^{-1}$ eV$^2$) | (eV) | | (×10$^{-3}$ μm$^2$) | | (×10$^{-3}$ μm$^2$) |
| T80Z15B5  | 32.8 | 31.8 | 4.7 | 0.788 | 9.2(2)×10$^3$ | 6.10(2) | 2.57(1) | 22.2(2) | 0.53(1) | 71.2(3) |
| T80Z10B10 | 31.4 | 30.5 | 4.3 | 0.774 | 9.0(2)×10$^3$ | 6.11(3) | 2.48(1) | 22.4(2) | 0.51(1) | 67.2(4) |
| T80Z5B15  | 31.3 | 31.4 | 4.4 | 0.785 | 9.6(1)×10$^3$ | 6.17(2) | 2.46(2) | 24.2(2) | 0.51(2) | 65.8(5) |
| T70Z25B5  | 30.0 | 29.2 | 4.4 | 0.774 | 9.8(3)×10$^3$ | 6.30(3) | 2.40(2) | 19.5(2) | 0.58(2) | 66.2(4) |
| T70Z15B15 | 27.5 | 26.4 | 4.0 | 0.818 | 7.4(2)×10$^3$ | 6.06(3) | 2.34(1) | 16.6(2) | 0.53(1) | 62.3(4) |
| T70Z5B25  | 25.7 | 25.4 | 3.4 | 0.771 | 7.8(2)×10$^3$ | 6.18(3) | 2.22(1) | 20.2(2) | 0.52(1) | 62.7(3) |
| T60Z35B5  | 27.2 | 26.5 | 4.1 | 0.806 | 9.1(3)×10$^3$ | 6.33(4) | 2.36(2) | 20.6(3) | 0.40(1) | 63.4(7) |
| T60Z20B20 | 25.6 | 24.4 | 3.9 | 0.837 | 8.9(2)×10$^3$ | 6.41(3) | 2.32(2) | 19.1(2) | 0.34(2) | 63.1(6) |
| T60Z5B35  | 22.0 | 20.6 | 3.3 | 0.832 | 6.1(2)×10$^3$ | 6.13(4) | 2.26(1) | 17.9(2) | 0.23(1) | 62.9(8) |

As was verified previously [11], tellurite glasses without rare-earth ion content belong to the group of diamagnetic glasses and thus the rotation angle $\theta$ is positive under an applied external magnetic field. Because of the diamagnetic nature of prepared glasses, spectral dependence of the Verdet constant can be approximated using the Sellmeier function using Eq.4a, where $A_M$ and $B_M$ are the Sellmeier fitting parameters. Obtained fits are shown in Fig.4 and both parameters, $A_M$ and $B_M$, are listed in Table 5.

(a) $$V = \frac{A_M E^2}{\left(B_M^2 - E^2\right)^2}$$
(b) $$n^2 = 1 + \frac{B_1 \lambda^2}{\lambda^2 - C_1} + \frac{B_2 \lambda^2}{\lambda^2 - C_2}$$

(4)

Based on the classical theory of electromagnetism proposed by Becquerel, the connection between optical dispersion of refractive index and Verdet constant of diamagnetic material $V_D$ can be also expressed as (Eq.5):

$$V_D = \frac{\gamma e}{2 m_e c} \lambda \frac{dn}{d\lambda}$$

(5)

where $e$ is the elementary charge, $m_e$ is the electron mass, $c$ is the speed of light and $\lambda$ is the corresponding light wavelength. This formula also contains multiplicative factor denoted as magneto-optic anomaly ($\gamma$) to quantify the level of agreement between experimental and calculated values[46]. As it is obvious, this relation is taking into account the spectral dispersion of refractive index $dn/d\lambda$ with the magnitude of Verdet constant and thus the high values of Verdet constants in dielectrics can be obtained only in spectral ranges close to the optical bandgap where the spectral dispersion increases rapidly due to onsetting optical absorption. The spectral dispersion of refractive index can be described in the region without optical absorption using, for example, the two-term Sellmeier model using Eq.4b, where $B_1$, $B_2$, $C_1$ and $C_2$ are the Sellmeier fitting parameters. Therefore, the measured refractive indices were fit using the two-term Sellmeier equation to obtain the optical dispersion term $dn/d\lambda$ in Eq. 5. and calculate the Verdet constant. The exact procedure is described in detail in Supplementary information section 4 and obtained Sellmeier parameters $B_1$, $C_1$, $B_2$ and $C_2$ are presented in Table 5. For better consistency between experimental and calculated values of the Verdet constant ($V_{cal}$) using the Eq.5, refractive index data for Sellmeier fit were used in the same wavelength range as the Faraday rotation measurements (see Fig. S7). The calculated Verdet constants at different wavelengths were then fit to the experimentally obtained values by least squares fitting method with free gamma parameter (Fig. S8) and are in good agreement with the experimental values which validates the use of the presented model. Obtained values of $\gamma$ parameter and calculated Verdet constant are listed in Table 5. It is also visible, that $\gamma$ is in range 0.771 to 0.832 and increases with higher content of ZnO+BaO.

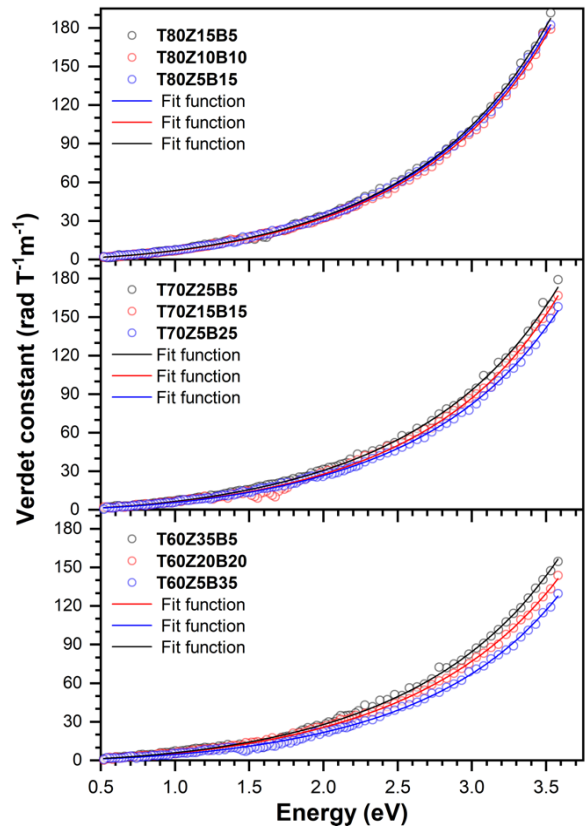

**Fig.4** Spectral dispersion of Verdet constant of TeO$_2$–ZnO–BaO glasses with indicated Sellmeier fit (parameters $A_M$, $B_M$ are listed in Table 5)

### 3.4. Optical fibers

To demonstrate the fiber-drawing ability of glasses from TZB system, composition 60TeO$_2$-20ZnO-20BaO was selected due to its good thermal stability ($\Delta T = T_x - T_c \geq 150$ °C), suitable temperature of glass transition ($T_g$ = 346° C), high refractive index ($n_{632}$ = 1.958) and its location approximately in the center of the glass formation region[21]. A cross-section view of the fabricated single-material fiber is shown in Fig. 5 with the observed diameter ≈150 μm and no visible defects. A cross-sectional view has been performed multiple times for different parts of fabricated fiber. Optical transmission of prepared fibers in NIR and MIR part of spectra was checked through cutbacks measurements, where the fiber exhibits characteristic spectral dispersion of optical absorption with an observed minimum loss of about ≈10 dB m$^{-1}$ in the range of wavelengths between 1550 and 1650 nm. This is a typical value when chemical purity of starting elements are not better than 4N (99.99%) [47]. Fabricated single-material fibers then allows satisfactory light propagation up to 2250 nm. Both minimum level of attenuation and transmission window can be significantly improved by using better quality starting products (i.e. with purity ≥5N) and through the involvement of purification procedures [48].

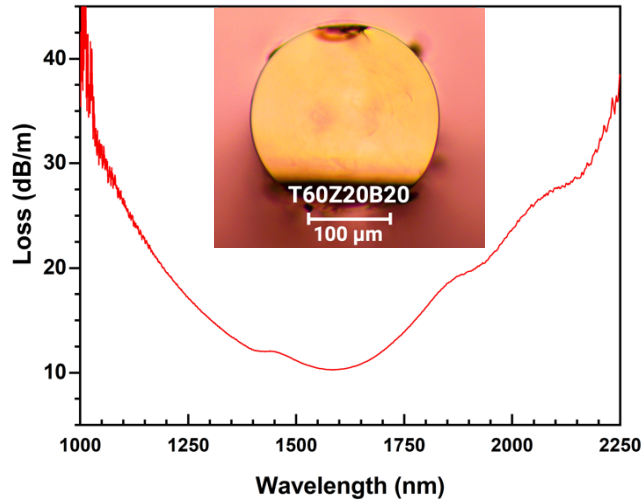

**Fig. 5** Optical attenuation curve and cross-section view of the multimode 60TeO$_2$–20ZnO–20BaO glass fiber.

## 4. Conclusions

A complex optical and MO analysis of nine selected glass compositions in the TeO$_2$–ZnO–BaO system has been performed with respect to the mutual substitution of each constituent element. Prepared glasses were investigated in wide spectral range $\lambda \approx 0.193$–25 μm for optical and $\lambda \approx 0.5$–2.3 μm for MO properties to determine the values of linear ($n_{632} \approx 1.91$–2.09) and non-linear refractive index ($n_2 \approx 1.2$–2.67×10$^{-11}$ esu) as well as the value of Verdet constant ($V_{632} \approx 22$–33 rad·T$^{-1}$·m$^{-1}$). From spectroscopic ellipsometry measurements were also estimated the values of optical band gap energy ($E_g \approx 3.73$–4.06 eV) and position of phononic/long-wavelength absorption edge using two new approaches. Obtained spectral dependence of refractive index was used to determine the parameters of dispersion energy ($E_d \approx 20.22$–21.82 eV), single oscillator energy ($E_0 \approx 6.88$–8.14 eV) and both first-order ($\chi^{(1)} \approx 0.198$–0.246) and third-order ($\chi^{(3)} \approx 2.59$–6.23 ×10$^{-13}$ esu) optical susceptibility. It was observed, that substitution of BaO by ZnO increases both linear and nonlinear refractive index, first-/third-order optical susceptibility and Verdet constant and decreases the value of optical band gap energy at constant TeO$_2$ content. Both linear refractive index and Verdet constant were used for estimation of magneto-optical anomaly parameter in diamagnetic materials for TZB system.

Optical fibers were manufactured from selected central composition 60TeO$_2$–20ZnO–20BaO and exhibit background attenuation level of ≈10 dB m$^{-1}$, which approaches the value of the tellurite glass fibers when prepared under room atmosphere conditions with no included purification process. Presented characterization will allow precise targeting of TeO$_2$–ZnO–BaO compositions with respect to the shaping processes and intended applications in photonics and fiber optics.

**ACKNOWLEDGEMENTS**

The work was supported by the Ministry of Education, Youth, and Sports of the Czech Republic, grant number LM2023037. The authors also thanks for the support by the European Regional Development Fund and the state budget of the Czech Republic (projects BIATRI: No. CZ.02.1.01/0.0/0.0/15\_003/0000445 0000445 and MATFUN CZ.02.1.01/0.0/0.0/15_003/0000487). J.H. acknowledges the support of the Charles University grant SVV–2023–260720 and project GA UK No. 662220.